\documentclass[prl,a4paper,twocolumn,preprintnumbers,nofootinbib,superscriptaddress]{revtex4}
\usepackage[a4paper, hdivide={1.91cm,,1.165cm}, vdivide={1.83cm,,2.5cm}]{geometry}

\usepackage{amsmath,amssymb}
\usepackage{graphicx}
\usepackage{units}
\usepackage[hyperfootnotes=false,colorlinks,citecolor=blue]{hyperref}

\newcommand{\beq}{\begin{equation}}
\newcommand{\eeq}{\end{equation}}
\newcommand{\bea}{\begin{eqnarray}}
\newcommand{\ena}{\end{eqnarray}}
\newcommand{\dd}{{\rm d}}

\allowdisplaybreaks

\begin{document}

\preprint{ULB-TH/19-05, UCI-TR-2019-17, \href{http://arxiv.org/abs/1906.10145}{arXiv:1906.10145} }

\title{Dark matter interactions with muons in neutron stars}

\author{Raghuveer Garani}
\email{rgaranir@ulb.ac.be}
\affiliation{Service de Physique Th\'eorique, Universit\'e Libre de Bruxelles, Boulevard du Triomphe, CP225, 1050 Brussels, Belgium}

\author{Julian Heeck}
\email{Julian.Heeck@uci.edu}
\affiliation{Department of Physics and Astronomy, University of California, Irvine, California 92697-4575, USA}

\hypersetup{
pdftitle={Dark matter interactions with muons in neutron stars},   
pdfauthor={Raghuveer Garani, Julian Heeck}
}


\begin{abstract}

Neutron stars contain a significant number of stable muons due to the large chemical potential and degenerate electrons. This makes them the unique vessel to capture muonphilic dark matter, which does not interact with other astrophysical objects, including Earth and its direct-detection experiments. The infalling dark matter can heat up the neutron star both kinetically and via annihilations, which is potentially observable with future infrared telescopes. New physics models for muonphilic dark matter can easily be motivated by, and connected to, existing anomalies in the muon sector, e.g., the anomalous magnetic moment or LHCb's recent hints for lepton-flavor non-universality in $B\to K\mu^+\mu^-$ decays. We study the implications for a model with dark matter charged under a local $U(1)_{L_\mu-L_\tau}$.
 
\end{abstract}

\maketitle


\section{Introduction}

Among the many odd properties of neutron stars (NS), one of the weirdest is certainly the existence of stable hyperon and muon layers. Much like the neutrons, these usually unstable particles are stabilized by the degenerate Fermi-gas phase, which Pauli-blocks the would-be daughter particles of a decay. 
A typical old NS contains roughly $10^{55}$ muons, compared to about $10^{57}$ neutrons~\cite{Bahcall:1965zza,Bahcall:1965zzb,Potekhin:2013qqa,Goriely:2013xba,Goriely:2010bm}. This lets NS interact with and potentially capture dark matter (DM) that couples mainly to muons, unique among astrophysical objects~\cite{Garani:2018kkd}.

Muonphilic DM might seem far-fetched, but considering our lengthy and so far unsuccessful quest for DM that interacts with first-generation particles in (in)direct detection experiments and at colliders it behooves us to consider alternatives to standard WIMPs~\cite{Bertone:2018xtm,Kopp:2009et,Kile:2014jea}. Furthermore, several anomalies that hint at physics beyond the Standard Model (SM) reside in the muon sector, e.g.~$(g-2)_\mu$~\cite{Bennett:2006fi,Keshavarzi:2018mgv} and $B\to K^{(*)}\mu^+\mu^-$~\cite{Aaij:2014ora,Aaij:2017vbb,Aaij:2019wad}, and typically require dominantly muonphilic interactions. A popular example here is the gauge boson $Z'$ of an anomaly-free $U(1)_{L_\mu-L_\tau}$~\cite{He:1990pn,Foot:1990mn,He:1991qd,Heeck:2011wj}, which can resolve either the $(g-2)_\mu$ anomaly~\cite{Gninenko:2001hx,Baek:2001kca,Carone:2013uh,Altmannshofer:2014pba} or the $B$-meson discrepancies~\cite{Altmannshofer:2014cfa,Crivellin:2015mga,Crivellin:2015lwa,Altmannshofer:2016jzy} depending on the $Z'$ mass $m_{Z'}$ and its gauge coupling $g'$. It is but a small step to connect such a new particle to DM, e.g., by using the $U(1)_{L_\mu-L_\tau}$ charge as a DM stabilization mechanism and the $Z'$ interactions to obtain the correct relic abundance. This results in dominantly \emph{muonphilic} DM, which can only interact non-gravitationally with NS.

Assuming a large enough muon--DM cross section to capture a significant number of DM particles, the infalling DM unavoidably transfers heat to the NS, see
 Refs.~\cite{Baryakhtar:2017dbj,Raj:2017wrv,Bell:2018pkk,Camargo:2019wou,Bell:2019pyc}.
This can increase the temperature of old NS from $\mathcal{O}(\unit[100]{K})$ to $\mathcal{O}(\unit[2000]{K})$, leading to an infrared blackbody spectrum that is in principle within range of future telescopes such as the James Webb Space Telescope, the Thirty Meter Telescope, or the European
Extremely Large Telescope~\cite{Baryakhtar:2017dbj}.
Further heating can occur through DM annihilations inside the NS.

In this paper we will discuss this capture of muonphilic DM in NS and the resulting heating, allowing for both DM and the mediator to be light to go beyond existing calculations. As a well-motivated example for muonphilic DM we consider an $U(1)_{L_\mu-L_\tau}$ model that can also ameliorate existing anomalies in the muon sector.

\section{Dark matter in neutron stars}
\label{sec:neutron_stars}

The capture of DM particles by the NS depends on the macroscopic NS properties such as its mass $M_\star$ and radius $R_\star$, and also on the properties of its Fermi-degenerate medium, especially chemical potential and lepton fractions. We consider a NS profile based on realistic two- and three-nucleon forces obtained in Refs.~\cite{Potekhin:2013qqa,Goriely:2013xba}, where phenomenological fits are performed for equations of state (EoS) of varying stiffness, neglecting the possibility of exotic quark phases. We present our results for the low-mass configuration of model BSK20 which corresponds to the NS values $M_\star = 1.52\,{\rm M_\odot}$, $R_\star=\unit[11.6]{km}$, the core muon fraction $Y_\mu=2\times 10^{-2}$, and muon chemical potential in the core $\mu_\mu =\unit[65]{MeV}$.
 
The above parameters for the NS are consistent with the observed properties of old NS. For example, number fractions of leptons in the core of a NS are important in the theory of NS cooling as they determine whether the direct Urca processes of neutrino emission operate efficiently or not~\cite{Klahn:2006ir}. Existence of beta equilibrium via the reactions $n \to p + e^- + \bar\nu_e$, $n \to p + \mu^- + \bar\nu_\mu$, and
$e^- \to \mu^- + \nu_e + \bar\nu_\mu$ together with charge neutrality ensures $\mu_n = \mu_p +\mu_e = \mu_p +\mu_\mu $ for the chemical potentials and $n_p =n_e +n_\mu$ for the number densities~\cite{Cohen1970}. Observations indicate that the direct Urca processes operate only in a relatively small number of NS, which could be explained by the so called {\it minimal cooling paradigm}, i.e.~not involving the direct Urca process~\cite{Page:2004fy,Pearson:2018tkr}. Moreover, the exceptions to the above paradigm can be explained by internal heating mechanisms~\cite{Potekhin:2015qsa} (none of which require DM inside the NS). All of the above indicate that an acceptable EoS should not allow the direct Urca process to occur in NS with masses below $\unit[1.5]{M_\odot}$~\cite{Klahn:2006ir}.
 
More recently, advanced versions of phenomenological EoS called BSK24 and BSK26~\cite{Pearson:2018tkr} were constructed by adding extra energy density functionals and by considering new atomic mass evaluation data. Since these models are fitted to the same EoS of moderate stiffness of Akmal--Pandharipande--Ravenhall~\cite{Akmal:1998cf} as that of BSK20, the qualitative predictions are very similar~\cite{Pearson:2018tkr}. Also note that these EoS are still allowed by the latest constraints from LIGO's observation of a binary NS merger~\cite{Abbott:2018exr,Most:2018hfd}. We refer to App.~A of Ref.~\cite{Garani:2018kkd} for more details regarding radial profiles of BSK20 and possible uncertainties on the DM capture rate.

\paragraph{Dark matter capture:}

\begin{figure}
  \includegraphics[width=0.49\textwidth]{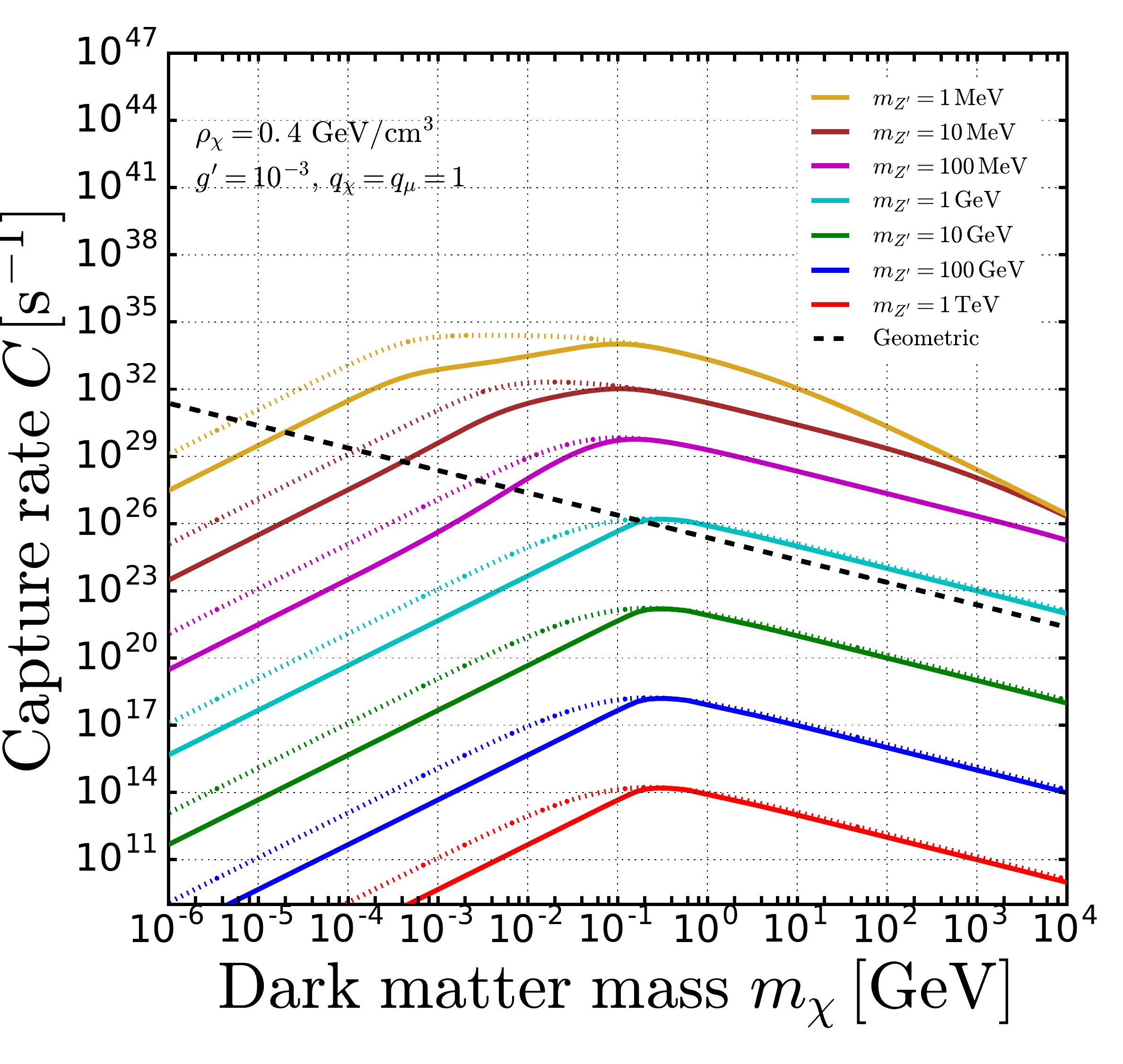}            
    
	\caption{DM capture rates on muons in NS for several $Z'$ masses (colored lines) as well as the maximal geometric capture rate $C_\star$ (black dashed line). The solid colored lines correspond to the exact capture rate~\cite{Garani:2018kkd}, the dotted  lines show the approximation from Eq.~\eqref{eq:cap-approx}.
	}
	\label{fig:cap-muons-med}
\end{figure}

The \emph{maximal} capture rate, i.e.~the rate at which all DM particles $\chi$ with mass $m_\chi$ that intercept the NS are captured, is given by the geometric rate~\cite{Garani:2018kkd}
\begin{align}
	C_\star \simeq \frac{5.6 \times 10^{25}}{\unit{s}} \, \frac{\rho_\chi}{\unit{GeV/cm^3}} \, \frac{\unit{GeV}}{m_\chi} \, \frac{R_\star}{\unit[11.6]{km} }\,\frac{ M_\star}{\rm 1.52\,M_\odot} \,,
	\label{eq:capturegeom-sim}
\end{align}
with DM energy density $\rho_\chi$ around the NS.
For DM--muon contact interactions this rate can be achieved for cross sections above $\pi R_\star^2/N_\mu \simeq \unit[5 \times 10^{-43}]{cm^2}$. 
The geometric rate $C_\star $ is shown in Fig.~\ref{fig:cap-muons-med} as a black dashed line fixing $\rho_\chi=\unit[0.4]{GeV/cm^3}$.

We will consider interactions between DM particles $\chi$ and muons that are mediated by a potentially light gauge boson $Z'$ with vector coupling $g' q_\mu$ to muons and coupling $g' q_\chi$ to $\chi$.
 The differential elastic scattering cross section in the non-relativistic limit is then 
\beq
 \frac{\dd \sigma }{\dd E_R}(\chi\mu \to\chi\mu) = \frac{ (g^\prime)^4 q_\chi^2 q_\mu^2 }{2 \pi}\frac{m_\mu}{w^2(2 m_\mu E_R + m_{Z^\prime}^2)^2}\, ,
 \eeq
valid both for scalar and fermion DM.
$w^2 = u_\chi^2 +v_\text{esc}^2$ is the squared speed of DM inside the NS, $E_R = m_\text{red}^2 w^2 (1- \cos \theta_\text{CM})/m_\mu$ is the recoil energy in the muon rest frame, and  $m_\text{red}$ the reduced mass of the DM--muon system.
The rate of accretion of DM particles from the halo into a NS by scattering off muons with density $n_\mu$ then reads~\cite{Feng:2016ijc}, in the zero-temperature limit,
\bea
C &\simeq&  \int_0^{R_\star} \dd r \, 4 \pi r^2 n_\mu(r) \int_0^\infty \dd u_\chi \, \left(\frac{\rho_\chi}{m_\chi}\right) \, f_{v_\star}(u_\chi) u_\chi \,   \nonumber \\
&& \times w(r)^2 \, \zeta(r) \int_{E_R^\text{min}}^{E_R^\text{max}} \dd E_R\, \frac{\dd \sigma}{\dd E_R} \,  .
\label{eq:cap-approx}
  \ena
To approximate Pauli blocking we introduce the factor $\zeta(r) \equiv {\rm Min}[1,\delta p(r)/p_\text{F}(r)]$ following Refs.~\cite{McDermott:2011jp,Bell:2013xk,Garani:2018kkd}.
Here $\delta p(r) \simeq \sqrt{2} m_\text{red} v_\text{esc}(r)$ is the typical momentum transfer in the muon rest frame and $p_\text{F}(r)=\sqrt{2 m_\mu \mu_\mu(r)}$ the Fermi momentum. 
$f_{v_\star}$ is the DM velocity distribution in the NS frame and $u_\chi$ the DM speed in the halo~\cite{Garani:2018kkd}. To capture a DM particle from the halo, the recoil energy should be between
 \begin{align}
 E_R^\text{min} = m_\chi u_\chi^2/2  && \text{ and } &&    E_R^\text{max} = 2 m_\text{red}^2 w^2/m_\mu \,.
 \end{align}
The $Z'$-mass dependence of the capture rate is given by
\begin{align}
\int_{E_R^\text{min}}^{E_R^\text{max}} &\dd E_R\, \frac{\dd \sigma}{\dd E_R}  = \frac{ (g^\prime)^4 q_\chi^2 q_\mu^2}{4 \pi w^2}  \, \Theta(4 w^2 m_\text{red}^2 - m_\mu m_\chi u_\chi^2)  \nonumber \\
& \times\frac{(4 w^2 m_\text{red}^2 - m_\mu m_\chi u_\chi^2)}{(m^2_{Z^\prime} + m_\mu m_\chi u_\chi^2) (m^2_{Z^\prime} + 4 m_\text{red}^2 w^2)} \,.
\end{align}
For $m_\chi \gg m_\mu > m_{Z^\prime}$ the capture cross section and rate thus become independent of the mediator mass.

For a close-by NS we take the local DM density $\rho_\chi=\unit[0.4]{GeV/cm^3}$ and present the DM capture rate on muons in Fig.~\ref{fig:cap-muons-med}. The colored dotted lines correspond to the capture rate obtained using Eq.~\eqref{eq:cap-approx}, whereas the solid curves are more accurate rates which are numerically obtained by using the methods of Ref.~\cite{Garani:2018kkd}, where possible non-trivial kinematics due to the degenerate nature of NS matter are taken into account.
For $m_{Z^\prime} >m_\chi > m_\mu$ the rate $C$ scales as $m_\chi^{-1}$; for $m_\chi \lesssim p_\text{F}/( \sqrt{2} v_\text{esc}) < m_{Z^\prime}$, Pauli blocking of the final state muon becomes efficient and the capture rate saturates and scales as $m_\chi^2$. For $m_{Z^\prime} < m_\mu$, Pauli blocking instead becomes important for $m_\chi \lesssim m_{Z^\prime}$. The full calculation of the capture rate differs from Eq.~(\ref{eq:cap-approx}) only when Pauli blocking is relevant and suppresses $C$ by a factor up to $ 35$. Thermal effects in the capture rate are negligible for the range of DM masses of interest here.
The main theoretical uncertainties in our calculation are the NS muon content and chemical potential, which should be at most off by a factor of two~\cite{Garani:2018kkd} (see also Ref.~\cite{Bell:2019pyc}).

\paragraph{Neutron star heating:}

As shown in Refs.~\cite{Baryakhtar:2017dbj,Raj:2017wrv,Bell:2018pkk,Camargo:2019wou,Bell:2019pyc}, the NS temperature increase due to DM capture is
\begin{align}
T_\text{\rm kin} \simeq \unit[1700]{K}\left(\frac{C}{C_\star} \right)^{1/4}\left(\frac{\rho_\text{DM}}{\unit[0.4]{GeV/cm^3}} \right)^{1/4} .
\end{align}
As can be appreciated from Fig.~\ref{fig:cap-muons-med}, it is not difficult to saturate $C\sim C_\star$ for much of the interesting parameter space. 
The SM--DM interaction underlying $C$ is of course impossible to reconstruct from $T_\text{\rm kin}$, but the observation of NS with $T < T_\text{\rm kin} $ in DM-rich environments would still allow us to set limits on DM--$\mu$ interactions. Importantly, truly muonphilic DM can \emph{only} be captured in NS, much like inelastic DM~\cite{Baryakhtar:2017dbj,Bell:2018pkk}, whereas most other DM models also allow for capture in other objects, including Earth.

Additional NS heating comes from annihilation of symmetric DM into SM particles. NS are expected to cool via neutrino emission through the modified Urca process for the first million years and through photon emission afterwards~\cite{Kouvaris:2010vv,Kouvaris:2007ay}. Once the rate of DM accretion equilibrates with DM annihilation the emissivity due to DM annihilation does not depend on temperature. This implies that once the temperature of the NS is sufficiently small, the power from DM annihilations that heat up the NS equals that of the photon emission, resulting in a constant temperature. For an old NS ten parsecs away, the maximal heating due to annihilations gives $T_\text{ann} \simeq \unit[2480]{K}\,[\rho_\text{DM}/(\unit[0.4]{GeV/cm^3})]^{0.45}$~\cite{Kouvaris:2007ay,Baryakhtar:2017dbj}.

Finally, comments on the impact of the NS profile on DM constraints from NS heating are in order. As discussed above the NS temperature increase due to capture and annihilation of DM are proportional to the capture rate, so uncertainties in the capture rate directly translate to the uncertainties on $T_{\rm kin}$ and $T_{\rm ann}$. In Ref.~\cite{Garani:2018kkd} it is found that the capture rate on muons differ by a factor $\simeq 5$ between the low-mass configuration of BSK20 ($\mu_\mu =\unit[65]{MeV}$, $N_\mu = 10^{55}$) and the high-mass configuration of BSK21 ($\mu_\mu =\unit[160]{MeV}$, $N_\mu = 3.8\times 10^{55}$). Note that $\mu_n$ and $N_\mu$  are not independent parameters. Consequently, the constraints obtained on the coupling will at most vary by a factor three. It is also important to note that uncertainties  due to the NS profile choice could be as large as the uncertainty on the local DM density.

For \emph{asymmetric} DM~\cite{Zurek:2013wia}, which cannot self-annihilate by construction, it seems possible to accumulate enough DM inside the NS to form a black hole that destroys its host~\cite{Goldman:1989nd}. The observation of very old NS in DM-rich regions would then allow us to put strong constraints on, e.g., the muon--DM cross section~\cite{Garani:2018kkd}. \emph{Fermionic} DM is typically difficult to constrain in this way due to the additional pressure from the Pauli exclusion principle. \emph{Bosonic} DM can in principle collapse much easier; however, as pointed out in Refs.~\cite{Kouvaris:2011fi,Bell:2013xk}, DM self-interactions play an important role and can hinder black hole formation even for bosonic DM, as discussed below.

\section{Dark matter charged under \texorpdfstring{$L_\mu-L_\tau$}{Lmu-Ltau} }
\label{sec:model}

\begin{figure}[tb]
	\includegraphics[width=0.48\textwidth]{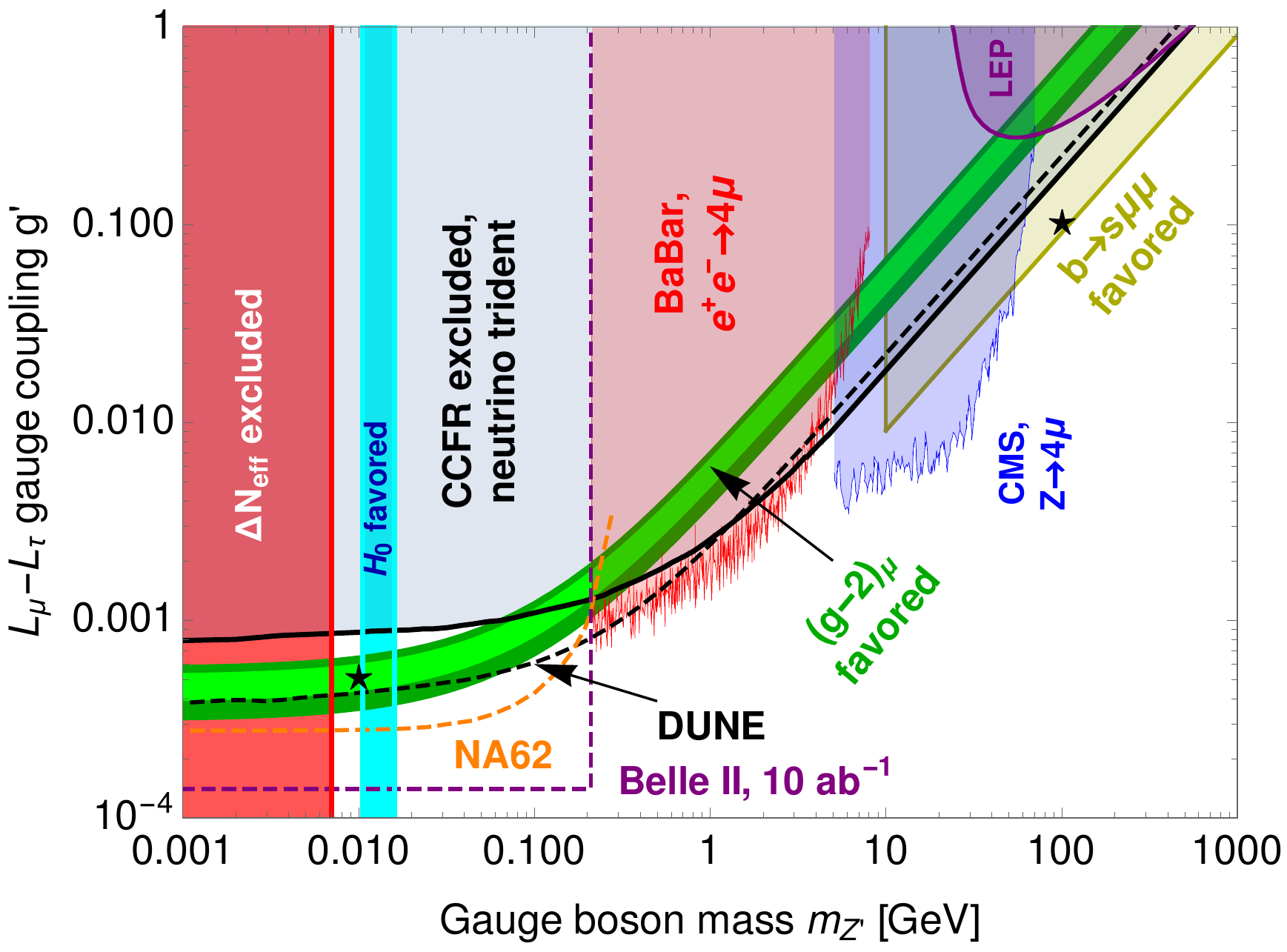}      
	\caption{
	Limits on the $U(1)_{L_\mu-L_\tau}$ gauge boson mass $m_{Z'}$ and coupling $g'$, assuming vanishing kinetic mixing. 
	The dark (light) green region can resolve the $(g-2)_\mu$ anomaly~\cite{Keshavarzi:2018mgv} at $2\sigma$ ($1\sigma$), the cyan region ameliorates the Hubble $H_0$ tension~\cite{Escudero:2019gzq}, and the yellow region can resolve the $b\to s\mu^+\mu^-$ anomaly~\cite{Altmannshofer:2016jzy} while satisfying $B_s$--$\bar{B}_s$ mixing constraints~\cite{DiLuzio:2017fdq}. 
	The other shaded regions are excluded by $N_\text{eff}$~\cite{Kamada:2015era,Kamada:2018zxi}, BaBar~\cite{TheBABAR:2016rlg}, CMS~\cite{Sirunyan:2018nnz}, LEP~\cite{Altmannshofer:2014cfa}, and neutrino trident production in CCFR~\cite{Mishra:1991bv,Altmannshofer:2014pba}.
	The dashed lines show the expected sensitivities of Belle-II~\cite{Jho:2019cxq}, DUNE~\cite{Altmannshofer:2019zhy,Ballett:2019xoj} and NA62 (in $K\to \mu +\text{inv}$)~\cite{Krnjaic:2019rsv}. Not shown are the sensitivities of M$^3$~\cite{Kahn:2018cqs} and NA64$\mu$~\cite{Gninenko:2014pea,Chen:2018vkr}.
	The two black stars denote the benchmark values used in the main text.
}
	\label{fig:Lmu-Ltau_limits}
\end{figure}

As a well-motivated example for muonphilic DM we consider an extension of the SM by the anomaly-free gauge group $U(1)_{L_\mu-L_\tau}$~\cite{He:1990pn,Foot:1990mn,He:1991qd,Heeck:2011wj}. Its gauge boson $Z'$ does not couple to first-generation particles and is thus only weakly constrained, as shown in Fig.~\ref{fig:Lmu-Ltau_limits}. It is arguably the simplest explanation of the $(g-2)_\mu$ anomaly if $g'\sim 5\times 10^{-4}$ and $M_{Z'} \sim 10$--$\unit[100]{MeV}$~\cite{Altmannshofer:2014pba}, fully testable with currently running experiments. For masses around $\unit[10]{MeV}$ this $Z'$ could furthermore ameliorate the observed tension in the Hubble parameter $H_0$ by contributing slightly to $N_\text{eff}$~\cite{Escudero:2019gzq}, and could also affect the high-energy neutrino flux measured in IceCube~\cite{Araki:2014ona,Araki:2015mya}. 
In a different region of parameter space, a heavy $L_\mu-L_\tau$ gauge boson could resolve the persistent anomalies in $b\to s\mu^+\mu^-$ transitions, as long as additional $Z_\mu'$ couplings to $\overline{b}\gamma^\mu P_L s$ are generated, e.g.~via vector-like fermions~\cite{Altmannshofer:2014cfa,Crivellin:2015mga,Crivellin:2015lwa,Altmannshofer:2016jzy}. This requires $m_{b}\ll m_{Z'}$ in order to obtain the desired contact-operator coefficient $C_9^\mu \simeq -0.95$~\cite{Aebischer:2019mlg} (see also~\cite{Alguero:2019ptt,Alok:2019ufo,Ciuchini:2019usw,Datta:2019zca,Kowalska:2019ley}) and $m_{Z'}/g' \lesssim \unit[1.1]{TeV}$ to evade the strong constraints from $B_s$--$\bar{B}_s$ mixing (at $2\sigma$)~\cite{DiLuzio:2017fdq}. In the following we will focus on two benchmark values for $L_\mu-L_\tau$ that satisfy the existing constraints from Fig.~\ref{fig:Lmu-Ltau_limits} and could play a role either for $(g-2)_\mu$ and $H_0$ ($m_{Z'}=\unit[10]{MeV}$, $g'=5\times 10^{-4}$) or $b\to s\mu^+\mu^-$ ($m_{Z'}=\unit[100]{GeV}$, $g'=0.1$).

\begin{figure*}[t] 
	\includegraphics[height=0.38\textwidth]{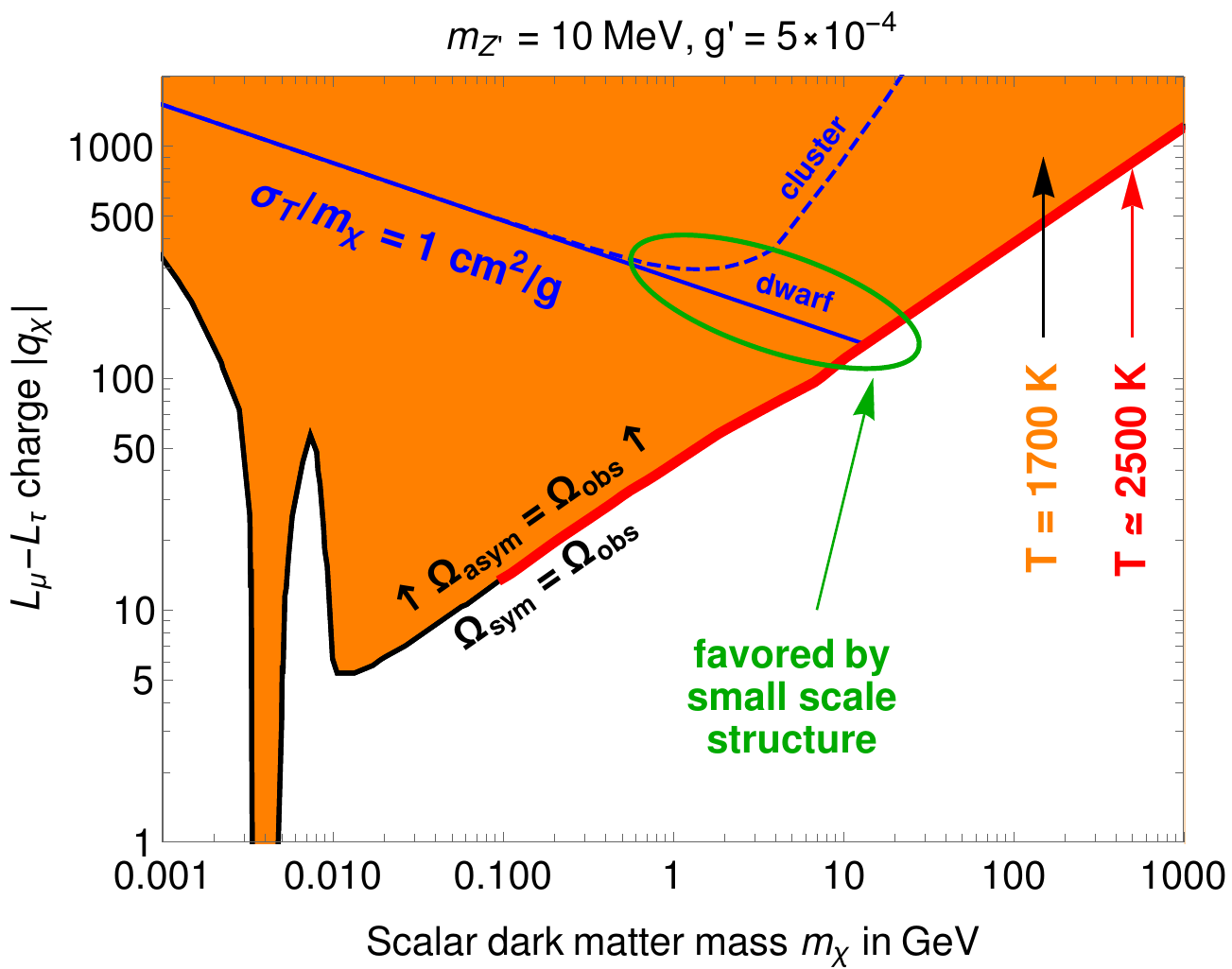}   
	\includegraphics[height=0.38\textwidth]{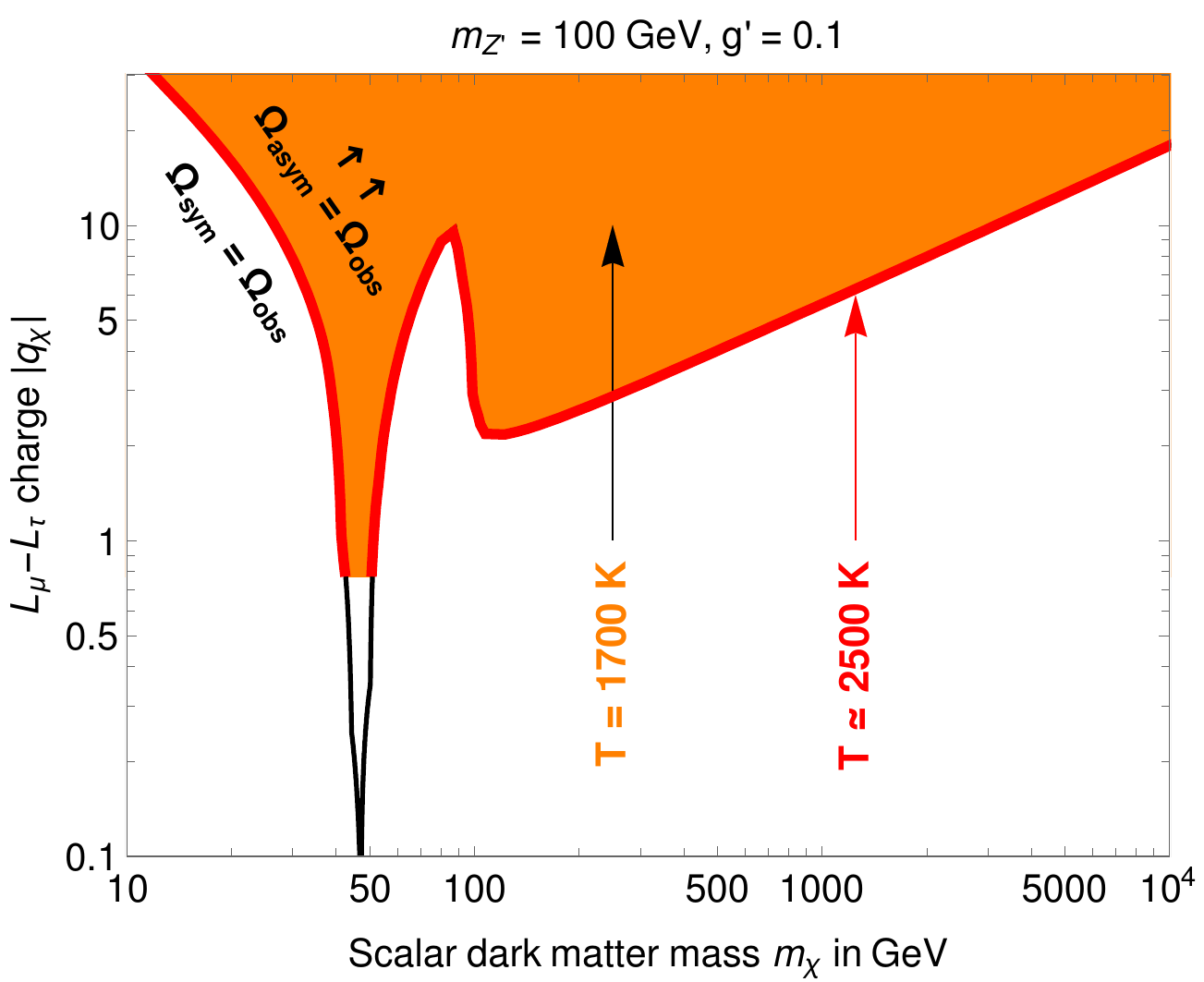}  
	\caption{
		Complex scalar $L_\mu-L_\tau$ DM $\chi$ for two benchmark values of $m_{Z'}$ and $g'$, one relevant for $(g-2)_\mu$ (left) and one for $b\to s\mu^+\mu^-$ (right). The black line denotes the WIMP scenario where the thermal symmetric abundance $\Omega_\text{sym}$ matches the observed DM abundance $\Omega_\text{obs}$; above is the region of asymmetric DM.
		In the orange parameter space we expect kinetic heating of old nearby NS to $\unit[1700]{K}$ by DM capture on muons; in the red region additional annihilation increases the NS temperature to $\unit[2500]{K}$.
		The blue line shows the DM self-interaction transfer cross section $\sigma_T (\chi\chi\to\chi\chi)/m_\chi = \unit[1]{cm^2/g}$ that can resolve small-scale structure issues~\cite{Kaplinghat:2015aga}, esp.~if suppressed on cluster scales as here for $ m_\chi\gtrsim\unit{GeV}$ (green ellipse).}
	\label{fig:abundance}
\end{figure*}

DM charged under $U(1)_{L_\mu-L_\tau}$ has been discussed for some time, typically as a WIMP with thermal abundance~\cite{Cirelli:2008pk,Baek:2008nz}. An appropriately chosen $L_\mu-L_\tau$ charge $q_\chi$ can render a new particle $\chi$ stable with \emph{muonphilic} $Z'$ connection to the SM -- the DM interactions with tauons being practically irrelevant. 
For most values of $q_\chi$ the full Lagrangian actually has an additional \emph{global} symmetry $U(1)_\chi$ that corresponds to conserved DM number. In analogy to the observed baryon number asymmetry one could assume that the cosmological history also led to a DM asymmetry~\cite{Zurek:2013wia} which results in a density $\Omega_\text{asym}$ proportional to the asymmetry.
Unless the $Z'$ couplings are tiny, DM $\chi$ will also be produced thermally, leading to a \emph{symmetric} DM component~$\Omega_\text{sym}$. We will consider both options here but remain agnostic about the potential DM asymmetry.
As a concrete example we will discuss \emph{scalar} DM, the fermionic case is similar.

For the two benchmark points specified earlier we show the scalar-DM parameter space in Fig.~\ref{fig:abundance}, obtained using {\sc MicrOMEGAs}~\cite{Belanger:2006is}, illustrating both the WIMP case, $\Omega_\text{sym}=\Omega_\text{obs}\simeq 0.26$, and the asymmetric DM region that arises for larger couplings.
The annihilation channels $\chi\bar\chi\to \mu\bar\mu$, $\tau\bar\tau$, $\nu\bar\nu$ are $p$-wave suppressed for scalar DM, whereas $\chi\bar\chi\to Z'Z'$ is $s$-wave and thus dominates the annihilation for $m_\chi>m_{Z'}$.
The orange regions indicate that the DM capture rate is maximized, $C\simeq C_*$, so a nearby NS would be kinetically heated to $\unit[1700]{K}$. $C\simeq C_*$ holds away from the resonance region $m_\chi \sim m_{Z'}/2$.	
Notice that our asymmetric scalar DM unavoidably has repulsive self-interactions mediated by the $Z'$. As a result, even asymmetric scalar DM will not lead to black hole formation inside the NS~\cite{Kouvaris:2011fi,Bell:2013xk}, unless the DM masses are far above the values considered here.

In the light-mediator case these repulsive self-interactions can also have an impact on structure formation, which has been argued to prefer DM--DM transfer cross sections $\sigma_T/m_\chi \sim\unit[1]{cm^2/g}$ on dwarf-galaxy scales (DM velocities $v_\text{dwarf}\sim\unit[10]{km/s}$)~\cite{Tulin:2017ara}. We show this contour in Fig.~\ref{fig:abundance} using the formulae of Refs.~\cite{Tulin:2012wi,Tulin:2013teo,Cyr-Racine:2015ihg} for $\sigma_T (\chi\chi\to\chi\chi)$. Cross sections of similar size do not seem to be favored on cluster scales, although the issue is currently far from settled. Suppressing $\sigma_T/m_\chi$ on cluster scales ($v_\text{cluster}>\unit[2000]{km/s}$) requires a velocity-dependent cross section~\cite{Kaplinghat:2015aga}, which arises in our model for $m_{Z'}/m_\chi \lesssim v_\text{cluster}/c$, as can be seen in Fig.~\ref{fig:abundance} (left) [see also Ref.~\cite{Kamada:2018zxi}]. 
The DM mass region $1$--$\unit[10]{GeV}$ in the light $Z'$ case is thus potentially preferred due to small-scale structure formation, while the region above the dashed blue line in Fig.~\ref{fig:abundance} is disfavored by observations such as the bullet cluster~\cite{Robertson:2016xjh}.

For \emph{symmetric} DM there is the possibility of additional NS heating through DM annihilations. Even the $p$-wave channels will reach equilibrium with capture~\cite{Kouvaris:2007ay,Kouvaris:2010vv} and lead to nearby NS temperatures up to $\unit[2480]{K}$~\cite{Baryakhtar:2017dbj}; only low-energy neutrinos ($E_\nu \lesssim \unit[0.1]{GeV}$) are able to escape the NS.
DM annihilations also give rise to indirect-detection signatures;
for the light $Z'$ case of Fig.~\ref{fig:abundance} (left) this is irrelevant due to $p$-wave suppression and difficult-to-observe $s$-wave neutrinos~\cite{Bringmann:2016din}.
For heavier $Z'$, the $s$-wave $\chi\bar\chi\to Z'Z'\to 4\ell$ could lead to indirect-detection signatures, but is currently only relevant for $m_{Z'}< m_\chi \lesssim \unit[100]{GeV}$~\cite{Leane:2018kjk}, i.e.~not for Fig.~\ref{fig:abundance} (right).

So far we have ignored additional interactions that arise from kinetic mixing or the Higgs portal, which will induce DM couplings to non-muonic matter and are highly constrained by direct-detection experiments~\cite{Altmannshofer:2016jzy,Araki:2017wyg,Arcadi:2018tly,Bauer:2018egk}.
For example, the one-loop scattering of DM via $Z'$ on the protons inside a nucleus $N$ takes the form~\cite{Kopp:2009et,Altmannshofer:2016jzy}
\begin{align}
\hspace{-1ex}\sigma_{\chi N} =\frac{Z^2}{A^2}\frac{m_{\text{red},\chi N}^2 }{\pi  m_{Z'}^4}\left(g' q_\chi\right)^2\left[e \epsilon+ \frac{ \alpha g'}{3 \pi } \log \left(\frac{m_\tau^2}{m_\mu^2}\right)\right]^2 ,
\label{eq:DM-proton}
\end{align}
keeping both the tree-level contribution from a Lagrangian term $\tfrac{\epsilon }{2}\, Z'_{\mu\nu} A^{\mu\nu}$~\cite{Galison:1983pa,Holdom:1985ag} and the finite one-loop contribution from muon and tauon loops~\cite{Holdom:1985ag,Araki:2017wyg}.
For heavy DM and $\epsilon=0$, XENON1T~\cite{Aprile:2018dbl} seemingly excludes the entire parameter space of Fig.~\ref{fig:abundance} (right) using Eq.~\eqref{eq:DM-proton}, while CRESST-III~\cite{Abdelhameed:2019hmk} excludes the entire $m_\chi \gtrsim \unit[1]{GeV}$ parameter space of Fig.~\ref{fig:abundance} (left). However, there is no reason to ignore the tree-level kinetic mixing angle $\epsilon$ and it is reasonable to expect other new particles in the model to contribute further to $\gamma$--$Z'$ mixing. As a result, the full kinetic mixing should be treated as a free parameter, constrained by direct-detection experiments. The observation of old cold NS can on the other hand set a constraint that is independent of the kinetic mixing angle and thus perfectly complementary to Earth's direct-detection experiments. 

This complementarity holds even more true for other muonphilic DM models. As a simple example, we can consider $U(1)_{L_\mu-L_\tau}$ with \emph{Majorana} DM $\chi$, which has a $Z'$ coupling $Z'_\rho\overline{\chi}\gamma^\rho\gamma_5\chi$ that leads to velocity-suppressed scattering cross sections. For the scattering on muons inside a NS this suppression is very mild because the infalling DM picks up a relativistic velocity, leading to a capture rate that is very similar to Fig.~\ref{fig:cap-muons-med}, at least for $m_\chi > m_\mu$. \emph{Direct detection} cross sections on Earth on the other hand will be heavily suppressed by $u_\chi^2 \sim 10^{-6}$ compared to the complex-scalar DM case of Eq.~\eqref{eq:DM-proton}, lowering the necessity to fine-tune the kinetic-mixing angle~$\epsilon$.
A similar disconnect between NS-capture and direct-detection cross sections can be achieved by replacing the $Z'$ mediator by, e.g., a pseudoscalar~\cite{Bell:2019pyc}.

\section{Conclusion}
\label{sec:conclusion}

The muon $g-2$ experiment at Fermilab (E989) is expected to confirm or dispute the longstanding magnetic-moment anomaly by the end of 2019~\cite{Keshavarzi:2019bjn}, while Belle~II will probe the $L_\mu-L_\tau$ solution to said anomaly. Together with LHCb, Belle~II is also expected to scrutinize the $B\to K^{(*)}\mu^+\mu^-$ anomalies that hint at lepton non-universality.
With these tantalizing and soon to be reevaluated hints for new physics in the muon sector it behooves us to consider muonphilic DM, which only has highly suppressed couplings to first-generation particles.
Using the curious fact that NS contain a large population of stable muons we can expect limits on DM--muon interactions from the DM capture on NS. 
Indeed, the infalling DM will heat old NS up to $\mathcal{O}(\unit[2000]{K})$ due to elastic scattering and potentially annihilation, which is in reach of future infrared telescopes.
For light mediator masses, relevant for $(g-2)_\mu$, muonphilic DM can also have self-interactions that resolve structure-formation issues.
Overall we have shown that muonphilic DM is not as elusive as naively expected, both in terms of motivation and in terms of signatures.

\section*{Acknowledgements}

We thank Mauro Valli for discussions and comments on the manuscript.
RG is supported by the ULB-ARC grant ``Probing DM with Neutrinos'' and the Excellence of Science grant (EOS) convention~30820817.
JH is supported, in part, by the National Science Foundation under Grant No.~PHY-1620638 and by a Feodor Lynen Research Fellowship of the Alexander von Humboldt Foundation. 

\bibliographystyle{utcaps_mod}
\bibliography{BIB}

\end{document}